# Twin-photon techniques for photo-detector calibration

**G. Brida, M. Genovese, M. Gramegna**
Istituto Elettrotecnico Nazionale Galileo Ferraris
Strada delle Cacce 91 – 10135 Torino, Italy

e-mail of corresponding author: genovese@ien.it

## Abstract

The aim of this review paper is to enlighten some recent progresses in quantum optical metrology in the part of quantum efficiency measurements of photo-detectors performed with bi-photon states.

The intrinsic correlated nature of entangled photons from Spontaneous Parametric Down Conversion phenomenon has opened wide horizons to a new approach for the absolute measurement of photo-detector quantum efficiency, outgoing the requirement for conventional standards of optical radiation; in particular the simultaneous feature of the creation of conjugated photons led to a well known technique of coincidence measurement, deeply understood and implemented for standard uses.

On the other hand, based on manipulation of entanglement developed for Quantum Information protocols implementations, a new method has been proposed for quantum efficiency measurement, exploiting polarisation entanglement in addition to energy-momentum and time ones, that is based on conditioned polarisation state manipulation.

In this review, after a general discussion on absolute photo-detector calibration, we compare these different methods, in order to give an accurate operational sketch of the absolute quantum efficiency measurement state of the art.

# 1. Introduction

In the recent years single photon detectors have found various important scientific and technological applications, which demand for a precise determination of their quantum efficiency. Among them the studies about foundations of quantum mechanics [1] , quantum cryptography [2], quantum computation, etc. Absolute calibration of photo-detectors requires the availability of sources of a known number of photons. Since at the moment solid state and atom in cavity sources are still too inefficient for these applications [3], spontaneous parametric down conversion (SPCD) heralded photon sources remains the only available system for absolute calibration.

Therefore, the outlook of this review is to report on the research frontier on the high accuracy calibration techniques for photon-detectors at single counting level.

Actually, in the realm of quantum optics, the photon entangled states give a unique and incisive means to measure absolute quantum efficiency of detectors since, because their intrinsic correlated nature, they supply the possibility to perform intrinsically absolute measurements based in principle only upon event counting, which are not tied to any other standard (absolute cryogenic radiometer, blackbody source or synchrotron emission).

A fundamental tool for this purpose is represented by bi-photon entangled states created by SPDC, an exclusively quantum effect arising from vacuum fluctuation firstly predicted by Luisell *et al.* [4] and independently investigated theoretically by Klyshko [5, 6], and that occurs in the interaction between a high energy electromagnetic field and the atoms of a nonlinear dielectric birefringent crystal (Fig. 1).

This spontaneous process manifests itself like a very low probable ($10^{-9}$) decay of a photon with frequency $\omega_{pump}$ into a twin conjugated photons of frequencies $\omega_{signal}$ and $\omega_{idler}$, obeying to constraint of energy and momentum conservation laws [7, 8], respectively:

$$\omega_{pump} = \omega_{signal} + \omega_{idler} \qquad (1)$$

$$\mathbf{K}_{pump} = \mathbf{K}_{signal} + \mathbf{K}_{idler}, \qquad (2)$$

where $\mathbf{K}_j$ are the relative wave-vectors; frequencies and propagation directions are determined by the orientation of the crystal, reflecting the so called phase-matching conditions, and ensuring an energy-momentum entanglement.

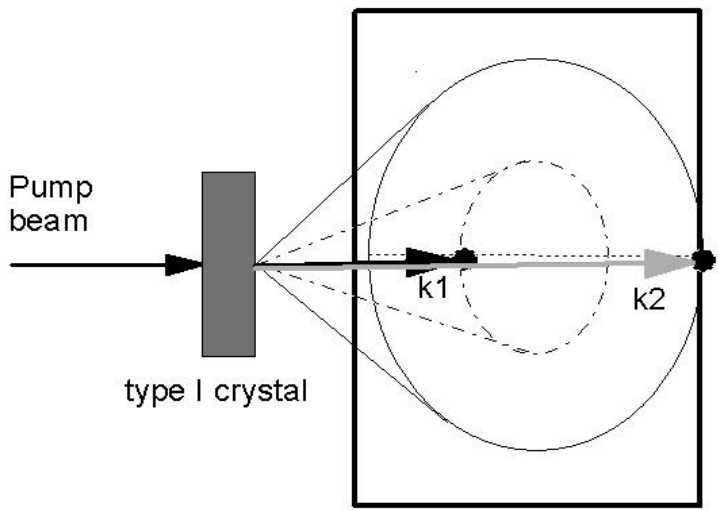

**Figure1.** Type-I Spontaneous Parametric Down Conversion.

The wave-function describing the pair of photons emerging from the decay are not expressed by a product of two well distinguishable states, and reveals the deeply nested nature of the two particles. At the first perturbative order of the expansion of the evolution operator it looks like [9]:

$$|\Psi\rangle \propto \frac{V d\omega}{2\pi} \sum_{\omega_1} \sum_{\omega_2} \phi(\omega_1, \omega_2) \frac{\sin \frac{1}{2}(\omega_1 + \omega_2 - \omega_3)t}{\frac{1}{2}(\omega_1 + \omega_2 - \omega_3)} \exp\left(i(\omega_1 + \omega_2 - \omega_3)t/2\right) |\omega\rangle_1 |\omega\rangle_2$$

(3)

where V is the pump laser amplitude, $\delta\omega$ is the mode spacing and $\phi(\omega_1, \omega_2)$ is a function which keeps into account geometrical factors associated to propagation of modes $|\omega\rangle_1$ and $|\omega\rangle_2$. As it is evident by Eq. 3, the two photons present a phase-momentum entanglement.

It must be mentioned that crystals showing second order susceptibilities [10] are transparent in the visible through the infrared frequencies, and are compulsorily birefringent to ensure phase-matching performances; this last feature is achieved in such crystals due to the fact that there are an ordinary and an extraordinary wave, which can have different propagation velocity and so keep a fixed phase relationship among them. In fact, it is possible to employ two different phase-matching schemes, the Type-I configuration that furnishes down-converted photons with parallel polarisation as ordinary waves, and the Type-II configuration [11, 12] with orthogonal polarisation conjugated photons.

Furthermore, the creation of the individual photon pair is broadband in energy, thereby the uncertainty on the emission time is very narrow, within the coherence time of the pump laser of tens of femto-seconds order.

This kind of simultaneous creation is, together with spatial correlation, the true basis for absolute quantum efficiency measurements, because it is intrinsic in the decay process that, revealing a photon on the signal propagation mode, the existence of the conjugated one in the idler branch is ensured, also with well defined direction, energy and polarisation; if the correlated one is not observed, it depends on the non-ideal efficiency of idler detector, that can be evaluated in such a way.

On the other hand, this characteristic has been well applied to carry out researches leaning to designing absolute photon sources [13], in the visible and infrared spectra, as well as to performing infrared radiation measurements without using infrared detectors and without absorption, and, finally, to characterize polarization dispersion parameters of optical and photonic components.

As shown in detail in the further sections, all the 'exotic' properties of correlated photons led to a well known technique of coincidence measurement, based on Type I SPDC, suggested and realised experimentally by Burnham and Weinberg and lately further independently elaborated theoretically by Klyshko: this system has been implemented in many laboratories (among which our) and deeply investigated to draft down an high-accurate uncertainty budget for metrological purposes and make it suitable for standard uses in national metrology institutes.

Otherwise, considering the currently expansion of research areas like quantum information, quantum communication and quantum cryptography, as well as the dream of a quantum computer, all fields in which is persisting the need for the chance of manipulating quantum states in high accurate way, at IEN, in collaboration with Moscow University, it has been performed an innovative set-up, that relies on a conditional polarisation rotation system, and that in principle not involves coincidence measurement. In such a scheme a polarisation correlated couple of photons is produced by type-II SPDC: if the detector on calibration branch clicks, a Pockels cell is activated on the optical path of the correlated propagation mode, acting on the polarisation state of the respective photon (90° rotation). The resulting degree of polarisation of the manipulated photon equals the quantum efficiency of the detector under test (DUT), giving an absolute measurement of quantum efficiency.

As it will be shown in the section 5, the two schemes are comparable, embodying at the same time different applicative perspectives.

## 2. Bi-photon states application fields outline.

The bi-photon states produced by SPDC applied to quantum efficiency calibration of photo-detectors, firstly employed experimentally by Burnham and Weinberg [14] and independently justified theoretically by Klyshko [15], were tested in many laboratories [16-20], albeit without an accuracy needed for metrology requests. The first true test of the method with independent verification was performed by J. Rarity's Research Group [21]; after some years at NIST [22] it was also realised a comparison between a conventional technique with a direct reference to an absolute standard and that one pursued by SPDC, showing an agreement between the two methods better than 2%. Later Brida et al. [23, 24] performed a careful calibration that pushed down to 0.5 % the correlated photon calibration uncertainty and to 1 % the conventional calibration one. Further recent developments can also be found in [25, 26]. The next goal is to reduce further systematic uncertainty down to 0.1 %, and this would constitute a noticeable improvement for infrared calibrations, to improve the calibration procedure in the region of low end optical power.

Before going deeper in the discussion about the absolute quantum efficiency calibration of single photo-detectors, it is interesting to have a look on what have been and are just now the other ways to employ the down-conversion phenomenon.

Exploring the possible applications furnished by conjugated photon states, it's easy to fall down into this consideration: when the working conditions on pump beam and on one down-converted mode are fixed, the constraint on simultaneous emission gives us information about observables describing the second conjugated photon, all without measuring, or better perturbing, the last one's state.

Otherwise, the pump radiation with the crystal and the trigger detector, all together, can be regarded as an absolute light source system in the visible and infrared spectrum region [27], that is very different from blackbody and synchrotron ones, because, at variance with them, it shares the knowledge of photon emission time, while in the classical source case only the average radiance is estimable. For sure, to implement this kind of a source, it's necessary take in account for non-ideal efficiency of trigger detector, false pulses from background noise of trigger device, missing photons in the output channel just announced by trigger (due to losses by reflection or absorption inside the crystal and in the optical path), events with two decays very close in time that furnish two photon from source versus a single count by trigger, and consequently optimise the arrangement.

The SPDC effect can also be amplified, as proposed by Klyshko [28] at Moscow State University and demonstrated by Penin's research group [29], to perform an evaluation of the absolute spectral radiance, namely the optical power per unit of area, per steradiant and per unit bandwidth, of an optical source, and also for high thermodynamic temperatures [30], without the need to tie to other reference standard.

In fact, SPDC is a non-resonant spontaneous phenomenon, with the proper feature that it can also be amplified by injecting a second source into the Non-Linear Crystal, saying, at the frequency of the idler photons $\omega_{idler}$, and matching for the spontaneous emission configuration: because all the output photons are created in pairs, the ones at the signal frequency, $\omega_{signal}$, are stimulated to be emitted and amplification is observed. This effect can be evaluated quantitatively and directly referred to the absolute quantum measurement, simply stopping the beam whose radiance has to be evaluated, and perform an acquisition of radiance only with the pump beam.

Further, the broadband emission permits to choose the working region of the spectrum, getting easier the calibration for infrared frequencies, without needing for an infrared detector, relying on trigger counts evaluation performed in the visible range. In fact the down-conversion process can be either frequency degenerate or frequency non-degenerate, and in the last case one photon can be in the infrared range while the twin in the visible spectrum.

With this method it is possible to work with monochromatic laser coherent light as well as with a broadband lamp radiation [29, 31]. On the other hand, the main difficulty in using this method for absolute radiance measurement of infrared light is the evaluation of mode coupling between the source under measurement and PDC emission [50].

Another metrological application of bi-photon's measurement of optical properties of materials that has been realized by L. Mandel's group [32] and Y. Shih's one [33]. The method relies on measures of polarization-mode dispersion, or difference in optical propagation times through a sample for light of orthogonal polarizations. The shape and the width of bi-photon wave function is measured by means of a nonlocal quantum-interference effect, pursuing a femtosecond resolution, that in more recent times has been improved and pushed-down to hundreds attoseconds at NIST [34].

Finally, correlated photon pairs were also used by Gisin's group for the measurement of chromatic dispersion in optical fibers [51].

## 3. Coincidence Measure Calibration

### 3.1 Measurement Scheme

In general the intrinsically absolute quantum efficiency calibration of single photon detectors is realized exploiting the Type-I SPDC arrangements, in particular the energy-momentum correlation and the simultaneous emission of the conjugated pair of photons.

The procedure consists in choosing the working frequencies of pump, signal and idler photons, to have a constrain in direction propagation modes, and placing a pair of photon counting detector modules to collect the flux of down-converted photons, (as shown in fig. 1): the firing of one detector, called trigger, warns the certain arrival of the conjugated photon, thanks to the above mentioned correlation properties. It's remarkable that for metrological purposes a certain angular dispersion for the single frequency due to finite crystal size and pump spectral bandwidth should be kept into account. Nevertheless, for usual values of crystal size and pump bandwidth, the correlations cones of the two photons of the pair are rather narrow, of the order of $10^{-4}$ radiants [35].

Let's assume N to be the total number of down-converted photon pairs emitted by the crystal in a set time interval $T_{gate}$, and $N_{signal}$, $N_{idler}$ and $N_{coincidence}$ are the mean number of events recorded, integrated on the same time interval $T_{gate}$, by signal detector, idler detector and in coincidence, respectively; so, we can write down the following obvious relationships:

$$N_{signal} = \eta_{signal} \cdot N \quad , \quad N_{idler} = \eta_{idler} \cdot N \qquad (4)$$

where $\eta_{signal}$ and $\eta_{idler}$ are the detection efficiencies on signal and idler arms.

Otherwise, the number of events in coincidence is

$$N_{coincidence} = \eta_{signal} \cdot \eta_{idler} \cdot N \qquad (5)$$

due to the statistical independence of the two detectors. Then the detection efficiency N signal follows:

$$\eta_{signal} = N_{coincidence} / N_{idler} \quad , \quad \eta_{idler} = N_{coincidence} / N_{signal} . \qquad (6)$$

It's trivial consider that to determine the efficiency of one detector, the signal one for example, it's useless the knowledge of trigger's quantum efficiency, because the non-ideal feature of this one doesn't affect in any manner the calibration of signal device; further, it isn't important to determine the transmittance of the optical elements for spectral selections placed in the trigger optical path.

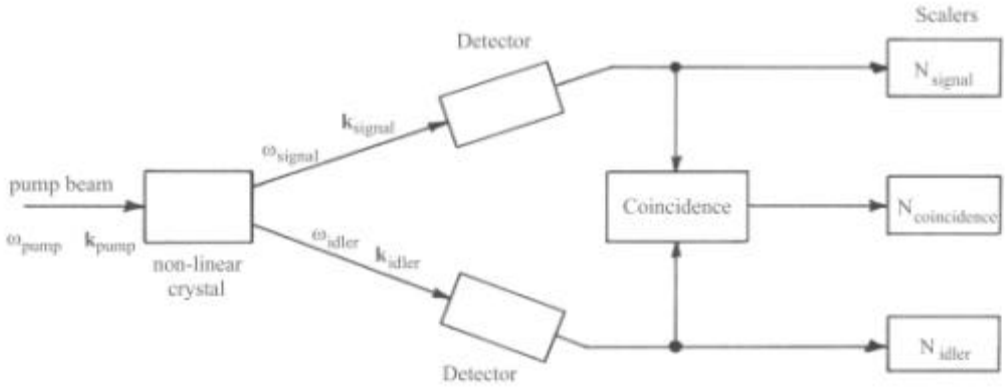

**Figure 2.** Coincidence Set-Up for Absolute quantum Efficiency Measurement.

### 3.2 Experimental set up description

As a detailed example of this technique [36], here we present one of the experimental realisations of the coincidence measurement method in our laboratory [37]. It consists of an Argon ion laser source, with 351 nm working wavelength, used to pump a 10 mm thick lithium iodate ($LiIO_3$) non-linear crystal (suitable for Type I SPDC phase-matching arrangement), housed in a sealed holder filled with an index-matching fluid. The working wavelengths for the down-converted is settled at 633 and 789 nm. The UV radiation downstream the crystal is absorbed by a filter.

It is useful to remark that estimate of $h_{signal}$ and $h_{idler}$ by equation (6) include the optical transmittance losses inside the crystal, as reflection, scattering and absorption effects, that are different for different frequencies: these losses can be measured by usual techniques with very high accuracy.

The spectral frequency selection on the correlated beam path is purchased in different ways on the two propagation directions: along the trigger direction a narrow wavelength selection is performed with a 3 nm FWHM bandwidth interferential filter centred at 789 nm, and spatial selection purchased by a 4 mm iris, all just before the collecting device; while on the other branch, the detector under calibration (DUT) avails a much broader wavelength selection, defined by the 8 mm diameter of the flux collector objective entrance, just to catch a fraction as large as possible of the correlated photons to the ones picked-up by the trigger one.

The detectors used in this experiment are single photon counting modules, in fiber-coupled arrangement (the measurement technique for a naked detector would be exactly the same) with a core diameter of 100 μm by means of lens. The down-converted light is collected and coupled to the fibers of each detector by means of 20x magnification microscope objectives (with 0.4 N.A.).

In the experimental implementation of the coincidence method the principal effects that can affect a high-accurate measurement are the transmittance $T_{signal}$ of the optical path on signal arm, the background contributions, deriving by diffused photons and detector dark counts, effects that must be estimated and corrected for the presence of spurious events and the loss of true coincidence events related to the acquisition systems.

In particular, in order to make a systematic comparison of various acquisition systems, the coincidence systems has been implemented in different ways, e.g. using two different Time to Amplitude

Converter (TAC), with and without valid start output, an high resolution time interval counter and an AND logical gate realised along the guidelines reported in [37].

For what concerns the TAC, the output pulses from the detectors are fed to the 'start' and 'stop' inputs of the converter which produces an output pulse whose height is linearly proportional to the time interval between the start and stop pulses. The TAC output is sent to a multichannel analyser (MCA) and to a single channel analyser (SCA). The MCA produces the time correlation histogram between start and stop pulse events, where the time correlated events produce a peak over a flat grass background originating by the uncorrelated events. The SCA generates a standardised amplitude output pulse for each input pulse with an amplitude included in a voltage window. By placing the upper and lower voltage thresholds of this window to select the events which fall in the coincidence peak of the MCA histogram, we obtain a countable pulse signal for the events in coincidence.

A delay line is inserted in the 'stop' channel in order to translate the peak in the histogram produced by correlated events of a time $t_{delay}$ from the origin of the time axis. Therefore, a true coincidence event, detected by both detectors, is missed if an uncorrelated 'stop' pulse arrives after the trigger 'start' event, at time t, but before the corresponding correlated one on the 'stop' channel, at time $t+t_{delay}$.

The fraction of missed coincidence counts is given by $\alpha \simeq 1 - N_{Dut} \, t_{delay} / T_g$ where $T_g$ is the gate time, the time interval of measurement.

A second counting system tested in our laboratory was a high resolution Time Interval Counter (TIC) with a resolution of 25 ps . The output pulses from the detectors are fed to the 'start' and 'stop' inputs of the counter and the measured time interval numerical values are sent to a computer. With this set-up instead of measuring over a time interval $T_{gate}$, we measure a defined number of 'start'-'stop' couples of events (ten thousands in our measurements, subdivided in 5 sub-samples) and we report in a histogram the number of measurements within a small time window (100 ps) vs. 'start'-'stop' time interval. The correlated events produce, like the TAC method, a peak; summing the number of events falling in the peak we have the number of coincidences $N_{coincidence}$. Concerning the missed coincidence counts correction $\alpha$, the same considerations exposed for the TAC remain valid. Finally, the AND gate does not show significative dead-time, so corrections are not necessary.

In conclusion a last correction must still be included. In fact, we have to account for the signal counts loss due to the detector dead time. In the simple situation of a fixed dead time, valid in principle for the avalanche detector with active quenching circuitry, we have a loss factor $\gamma$, analogous to the previous correction terms.

Taking into account all the correction terms reported above, the final equation for the detector quantum efficiency is:

$$\eta_{SIGNAL} = \frac{1}{T_{SIGNAL}} \cdot \frac{N_{coincidence} - N_{accidental}}{N_{trigger} - N_{background}} \cdot \frac{\beta}{\alpha \gamma} \qquad (7)$$

where $T_{SIGNAL}$ is the transmittance of the signal photon path, $N_{accidental}$ and $N_{background}$ can be respectively estimated measuring coincidence rate far from the peak,
and turning–off the parametric fluorescence light by a 90° rotation of the crystal (or of the half-wave plate), and counting the number of events on trigger ($\beta$ =1 except for a TAC without valid start output signal, where it keeps into account overestimation of trigger counts).

It's remarkable that a possible development of this calibration technique would be by use of a pulsed laser source, to pursue high peak power of the pump beam with a corresponding increase in the probability of parametric down conversion. The gating capability of the detectors modules (if present for the detector under measurement) with a signal sinchronized with the laser pulse emission would allow the inhibition of the detector without pump signal, eliminating the detectors dead time problems. The repetition rate is limited by the gate turn-on turn-off delay to a maximum value of the order of the inverse of these time delays.

In order to select precisely the maximum of the quantum efficiency a precise scan in both axes must be performed. In Fig. 3 it is presented one section of this scansion realised for calibration of a APD detector at 633 nm. By the collection of various scansions of this kind, and by the use of a suited collection optics before the detector, one can eventually reconstruct spatial response of the photo-detector.

Finally, two variation on this theme are, firstly, a calibration method using only one photo-detector, with one photon of the pair opportunely delayed, and performing an auto-correlation evaluation of the counts [38] using PMT able to detect multiphoton states.

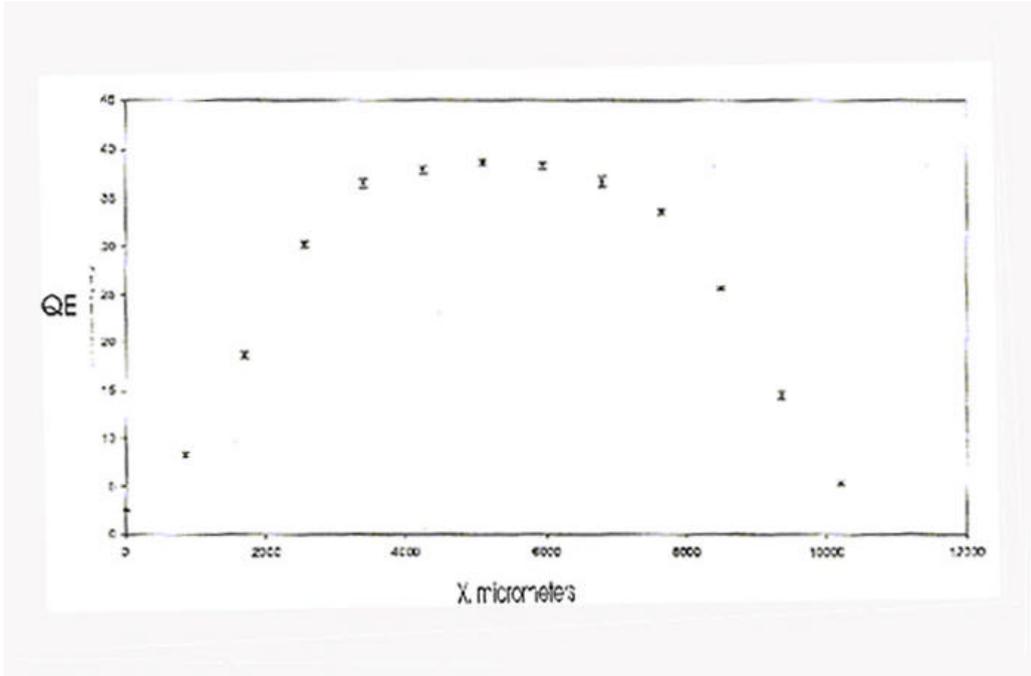

**Figure 3** Experimental Quantum Efficiency curve in function of the detector position X (with TAC Coincidence circuit acquisition).

## 4. Conditional Polarization Rotation technique Calibration

### 4.1 SPDC Source and Measurement Scheme

As a second scheme of absolute calibration of single photon detectors we present an innovative, very recent measurement technique realized with a Type II SPDC [39, 40] source of correlated pairs (Fig. 4) in contrast to Type I phase matching set up for coincidence measurement, to take advantage of orthogonal polarisation of the down-converted photons. In fact, in Type-II case when occurs a decay inside the crystal, if the signal photon is vertically polarised, the idler one would be horizontally polarized, and vice versa, realizing a polarization-entangled state:

$$|\Psi\rangle = \frac{|HV\rangle + e^{if}|VH\rangle}{\sqrt{2}} \tag{8}$$

where H and V respectively denote horizontal and vertical polarization state of photons, and **f** can be varied by small tilt of the crystal or inserting birefringent elements. The schematic of this new set-up is reported in fig.5.

One photon of the pair is directed, after a polarization selection, to the detector D1 under calibration (Fig. 5). When this photon (signal) is detected, on the correlated arm the polarization of the delayed (by means of optical fibre F) second photon (idler) of the pair is rotated of 90° by means of a fast high-voltage switch, the Pockels cell (PC). In this way the polarization state of the second photon strictly depends on the quantum efficiency $h_1$ of the detector D1, since the Pockels cell is activated only when the first photon is effectively detected. The count rate $W_2$ of the detector D2 is given from the following expression:

$$W_2(\theta) = \frac{\alpha \cdot \eta_2 \cdot W_0}{2} \cdot (1 - \eta_1 \cdot \cos(2\theta)) \qquad (9)$$

where $\alpha$ is the idler optical path transmittance (fibre and Pockels cell transmittance factor), $\eta_2$ is the D2 quantum efficiency, $W_0$ the rate of photon pairs and the angle $\theta$ is the polarizer (G) setting, the angle $\theta = 0°$ corresponding to horizontal transmission. The visibility V of the count rate signal $W_2$ obtained by rotating the polarizer preceding the detector D2 is a measurement of $\eta_1$:

$$V = \frac{\max(W_2) - \min(W_2)}{\max(W_2) + \min(W_2)} = \eta_1 \qquad (10)$$

It is then realised a measurement of absolute quantum efficiency without needing an evaluation of coincidence counts.
It's important to point out that this kind of trigger conditional polarisation manipulation on correlated photons works in real-time.

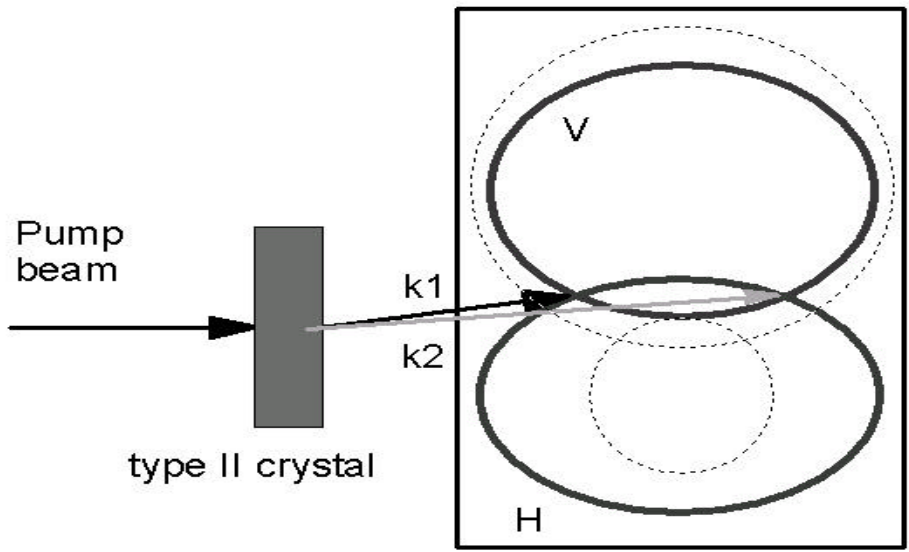

**Figure 4.** Type-II Spontaneous Parametric Down Conversion

### 4.2 Experimental set up

The experimental set-up arrangement in our laboratory [41, 42] include a CW Argon ion laser at the wavelength of 351.1 nm that irradiate a non-linear BBO crystal (5 mm side cube) in type-II SPDC phase-matching. For sake of completeness it's remarkable that no compensator has been inserted after the crystal, setting the phase at the mercy of classical random fluctuation, and giving mixed state, actually not really polarization entangled, but with no invalidation on the results for these purposes.
The residual pump beam after passing through the BBO crystal is absorbed by a beam dumper.
We selected from the output broadband parametric fluorescence emission the directions of correlated, frequency degenerate, photon pairs with 702 nm wavelength. After crossing a polarizing beam splitter (PBS) and selecting vertical polarization, the first (signal) photon of the pair was detected by means of a silicon avalanche single photon detector D1 preceded by a pinhole and a red glass filter (RG) (which together constitute the detection apparatus to be calibrated). The optical path of the conjugate photon (idler) of the pair has been lenghtened by means of a coupling in a 50 m single mode (4 μm core diameter) polarization maintaining fibre (F), to delay its arrivals. Input and output fibre couplings are realized with 20X microscope objectives (0.4 NA).

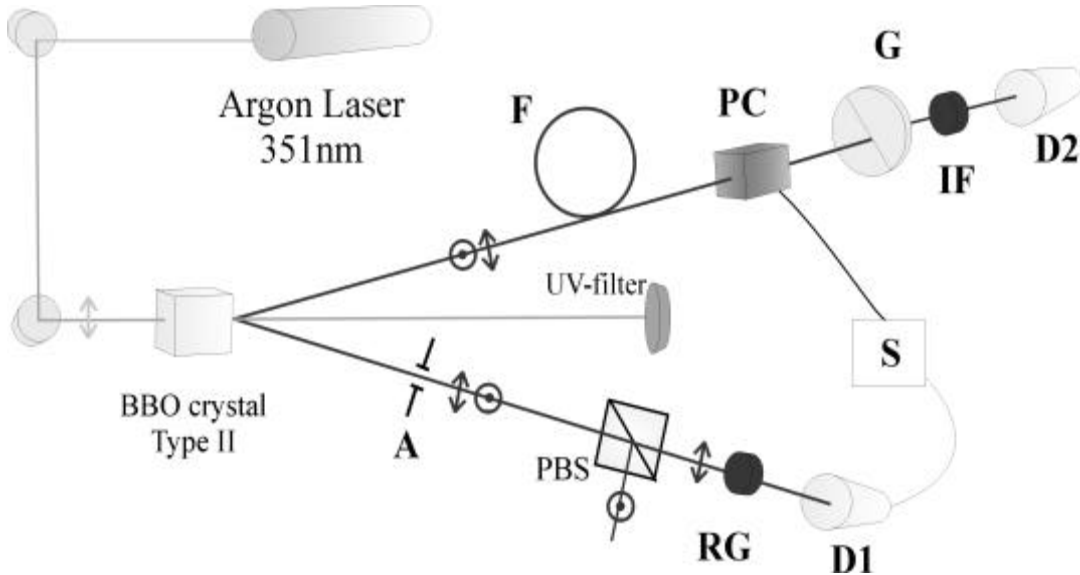

**Figure 5.** Conditional Polarization Rotation technique for Absolute Quantum Efficiency Calibration

The idler photon was then directed to a KDP Pockels cell followed by a Glan-Thompson polarizer (G), an interference filter (IF) at 702 nm (4 nm FWHM) and a second silicon avalanche single photon detector D2. When the Pockels cell driver (S), was triggered by the detector under calibration, D1, it generated a high-voltage pulse (5.2 kV) with fast rising edge (5 ns), a 180 ns flat-top and a long fall tail of 10 μs duration. The Pockels cell was operating as a half-wave plate oriented at 45° to the vertical axis. The Pockels cell controlled in this way realizes a 90° rotation on the polarization of the second photon conditioned to a measurement of a vertically polarized photon on the conjugated arm. To ensure that the flat part of the high-voltage pulse arrives at the Pockels cell simultaneously with the signal photon delayed by the optical fibre a fine, variable, electronic delay adjustment was added.

The figure 6 shows the trend of the counts measured by detector D2 for an integration time of 10 s (without background subtraction): in the case of absence of any trigger signal by detector D1 the behaviour versus the angle orientation of the polarizer is nearly flat (squared symbol), no preference in the state of polarization is expected for the photon pairs generated by type-II SPDC. The small fluctuation of this curve is likely related to a residual coupling misalignment of the single mode polarization maintaining fibre with respect to the horizontal and vertical axes. In the case of conditioned polarization rotation control (triangle symbol), the experimental points spread on a sinusoidal modulation function dependent on the polarizer orientation. In this case the Pockels cell effect is to flip an horizontal polarized photon into a vertical one whenever a vertical polarized photon is detected by D1.

As seen in equation (10), the visibility on the count rate signal $W_2$ obtained by rotating the polarizer G preceding the detector D2 is a measurement of the quantum efficiency of detector D1, and aiming to an high accurate measurement of this parameter, it is fundamental have a look to what can affect the measurements. To do so it's first of all better to obtain a background estimation (by 90° rotation of the pump laser polarization) and subtract this amount to the visibility counts: the visibility can be evaluated from the maximum and minimum value of the recorded data.

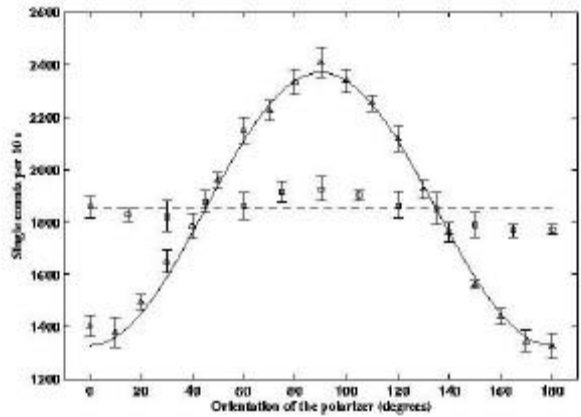

**Figure 6**. Counts measured by detector D2 for an integration time of 10 s: flat squared symbols trend means the absence of trigger signals, while the triangle shaped symbols present the typical modulation performed by the triggered Pockels cell.

The determination of the non-ideal efficiency of the Pockels cell can be performed with a simultaneous acquisition of counts with a coincidence system, as we done in view of the comparison of the two methods.
Incidentally, the visibility obtained shows a large uncertainty due to the poor estimate of the count rate minimum which is comparable with its fluctuations (detector dark counts, background scattered light): for this reason the experimental results were treated with a least square adjustment (LSA) approach [43].
Moreover, the other correction items are for the electrical dead times of D1 and Pockels cell driver, and also for the optical losses in the polarizer cube in the signal path. The dead time correction is largely dominated by the maximum working rate of the Pockels cell driver, that must be kept under 10000 counts rate per second. The fixed dead-time is 10 μs.
As stated before, the quantum efficiency evaluated in this way is not the "naked" detector one, but the one corresponding to the detection apparatus including spatial and spectral filtering.
To pursue the efficiency of the bare detector would be necessary to introduce a corrective factor keeping into account losses in the other elements in the detection apparatus (red glass filter) and in the non-linear crystal (corrections needed of course also for the calibration method based on correlated photon).

### 5. Measurements Comparison

To check the robustness of the conditional polarisation rotation method, simultaneously with the acquisition of detectors counts, the output signal were split and redirected also to a connected coincidence acquisition system.
The auxiliary experimental data of coincidence events (between D1 and D2) performed with and without the conditioned polarization rotation are presented in fig. 7, in which is well shown the half-wave plate like behaviour of the system from the 90° shift between the two curves.
The results for both the calibration methods have been corrected for systematics following the guide-line considerations mentioned above, and present an absolute quantum efficiency evaluation value for coincidence set-up of $h_1 = 0.486 \pm 0.002$, against a value for the innovative arrangement of $h_1 = 0.486 \pm 0.011$ [44].

These results are compatible and comparable, and represent a clear proof of the competitiveness of the new method.

The bigger value for the conditional polarisation rotation arrangement means that a defintive uncertainty budget has not been compiled yet: a substantial reduction can be expected by a further deep study of the measurement systematics.

The new calibration scheme is more elaborate than the traditional calibration, because it requires a real-time feed-forward control of the polarization state of a photon.

One potential attractive of the new scheme is the possibility to extend the method to the calibration of analog detectors.

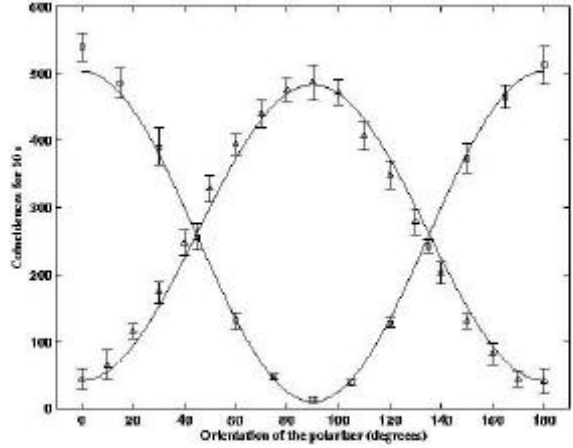

**Figure 7**. Experimental data of coincidence events performed with and without the conditioned polarization rotation.

**6. Outlooks for absolute calibration of analogical photo-detectors**

Finally, we would like to hint to possible extension of PDC calibration methods to absolute calibration of analog detectors.

A first attempt in this sense was presented in [45]. The method is based on the circumstance that the light flux can be expressed as the sum of random train of pulses. By simultaneously measuring the autocorrelation function of the currents of each of the detectors and the mutual correlation function, the quantum efficiency of one of them $(\eta)$ can be expressed as the ratio of the mutual correlation function $<i_1 i_j>$ over the autocorrelation $<i^2>$ of the other detector multiplied for a corrective factor $K(i,j)$ keeping into account the fluctuations of the gain and other noise characteristics of photo-detectors:

$$\eta_2 = K(1,2) < i_1 i_2 > / < i_1^2 > \qquad (11)$$

Nevertheless, this method is applicable only in the limit of low intensities [46].

More recently, expressions involving the quantum efficiency of analog detector (and therefore in principle usable for extracting it) were derived in studies concerning sub-poissonian light [47-49]. However, this initial results were not elaborated in the sense of really developing a calibration scheme.

A work specifically addressed to a metrological study of calibration of analog detectors is therefore of the utmost interest and now under realisation [46].

## 7. Conclusions

In this review we depicted an outline of possible implementation of bi-photon states from SPDC effect to metrology, in particular to describe the new challenges of an innovative method for absolute quantum efficiency calibration of photodetectors based on a conditioned manipulation of polarization states, in comparison with the well known coincidence measurement system, both techniques intrinsically absolute, not tied to any other radiometric standard. The experimental results are comparable and shows that the new method is competitive and suitable for further deep investigations, and gives also a chance to extend the method for absolute calibration of analog detectors.

**ACKNOWLEDGMENT**

This work has been supported by MIUR (FIRB RBAU01L5AZ-002 ), by "Regione Piemonte", Intas (grant #01-2122) and by Fondazione San Paolo.
We would like also to acknowledge that the experiment on Conditional Polarization Rotation technique Calibration has been performed in collaboration with M. Chekhova, L. Krivitsky and S. Kulik of Moscow University and M.L. Rastello of IEN, which we thank for very valuable collaboration and pleasure of working together.

Corresponding Author:

Marco Genovese
Tel: 39 011 3919253
Fax: 39 011 3919259

e-mail: genovese@ien.it